\documentclass[twocolumn,prl,showpacs]{revtex4}

\usepackage{latexsym}
\usepackage{hyperref}
\usepackage{epsfig}
\usepackage{graphics}
\usepackage{amsmath}
\usepackage{xspace}
\usepackage{amssymb}
\usepackage{amsthm}
\usepackage{dcolumn}

\def\be{\begin{eqnarray}}
\def\ee{\end{eqnarray}}
\def\nn{\nonumber}

\begin{document}

Submitted to Physical Review Letters and the WWW on \today.
\title{Gauge Theory of Pairing and Spin Fluctuations Near the Quantum Critical Point}

\author{J. R. Schrieffer}
\affiliation{NHMFL and Department of Physics, Florida State
University, Tallahassee, FL 32310}

\begin{abstract}
We develop a new theory of pairing and magnetic spin fluctuation
effect near the quantum critical point. Several novel properties
are predicted: 1) based on a spin fermion model, we derive two new
interactions, a) a spin deformational potential $H_{sdp}$
proportional to the bandwidth $W$ (as opposed to the considerably
smaller exchange coupling $J$ of the nearly antiferromagnetic
Fermi liquid theory) and b) a diamagnetic potential $H_{dia}$,
quadratic in a gauge potential $\vec{A}$. A dramatic increase of
$T_C$ is predicted for $0.01\: W\leq J\leq 10\:W$. This should
have immense technological impact in electric energy production,
storage and transmission, as well as for medical electronics,
microwave electronics, computer memory and information storage,
separations technology and maglev, amongst others. The striking
prediction to be confirmed by experiment is that the pairing order
parameter $\Delta(\vec{k})$ is predicted to be $p-$wave, i.e.,
$l=1$, $S=1$, as compared to $l=2$ and $S=1$ for conventional HTS
materials. In addition a novel collective model is predicted whose
frequency, $\omega_L$ is in the optical range and is determined by
$H_{sdp}$.

\end{abstract}

\pacs{71.10.Ay, 71.10.Pm.}

\maketitle

The static magnetic susceptibility of the localized spins in
approximately $30$ doped intermetallic compounds is found to be
highly anomalous namely,
\be \chi(\vec{Q},\omega_n=0,T)\propto1/T^\gamma,\quad
\textnormal{with}\:\gamma\simeq0.14\ee
where $\vec{Q}$ is the magnetic spin ordering vector of the spin
ordered phase, e.g. $\vec{Q}=(\pi/a,\pi/a)$ for a cubic
commensurate antiferromagnet. Equally anomalous is the specific
heat
\be C_V(T)\propto\ln T.\ee
This is to be contrasted with
\be\chi({Q,\omega_n=0,T})\propto\frac{1}{T-T_N}\quad\quad
(Ne\acute{e}l\:law) \ee
and
\be C_V(T)\propto T \nn \quad\quad\quad\quad\quad(Fermi\: liquid\:
law)\ee
of mean field theory. Anomalously small critical exponents are
also found in some ten other experiments. Thus, a universal
experimental behavior is found, i.e., a law of corresponding
states is observed, calling for a universal theory producing these
anomalous exponents. Such a theory is developed here, for the
first time.

In the pioneering work of Hertz\cite{Htz}, ferromagnetic critical
spin fluctuations were treated using a one fermion loop effective
action for the spin-fermion model. He found highly anomalous
properties, as in experiment. This work was extended by
Millis\cite{Mil} to the several fermion loop level. Since then, a
great deal of effort\cite{Chu}-\cite{Lch} has been expended with
little progress having been achieved. Here we give a complete
solution to the problem, including pairing, for the spin-fermion
problem and find $1) $excellent agreement with experiment for the
anomalous critical exponents, and $2)$ we predict a highly novel
$p-$wave, $S=1$ superconductor, a very high frequency Leggett mode
and an extraordinarily high $T_C$ of immense technological and
scientific interest.

For clarity, we use the spin fermion model.
\be H&=&H_0+H_J
\label{1}\\
H_0&=&-\sum_{i j s}t_{i j}\psi_{i s}^\dag\psi_{j s}\label{2}\\
H_J &=& J\sum_{i s s'}\psi_{i s}^\dag\psi_{i s'}\vec{\sigma}_{s
s'}\cdot\vec{S_i}\label{3} \ee
where,
\be \{\psi_{i s}^\dag,\psi_{j s'}\}&=& \delta_{i j}\delta_{s
s'}\label{4} \ee
\be [S_{i \alpha}, S_{j \beta}]&=& i \hbar S_\gamma \delta_{i
j}\label{5} \ee
with $\alpha, \beta, \gamma=1,2,3$ in cyclic permutation. Equally
well we could have use the Hubbard model with equivalent results,
although the mathematics would be considerably more involved.

A key aspect of the quantum critical point problem is that all
fluctuations, pairing and spin, are of long range ($L\gg a$,
$a$=lattice spacing), and of low frequency compared to the
electronic band width $W$ and the two site Heisenberg exchange
interaction, $J_2\simeq J^2/W$. This suggests that we develop a
long wavelength, adiabatic approximation. This we do below.


To exploit these slow variations, we introduce the unitary
electron spin rotation operator,
\be U(t)&=& T e^{\frac{i}{2}\sum_{i s s'} \psi_{i s}^\dag (t)
\vec{\sigma}_{s s'}\cdot\psi_{i s'}(t)}{}\\\nn & &
{}\sin^{-1}z\times
S_i(t)\:\frac{\hat{z}\times\vec{S}_i(t)}{|\hat{z}\times\vec{S}_i(t)|}.
\label{6}\ee
$T$ is the Feynman time ordered product defined by
\be T [A(t_1)B(t_2)C(t_3) \dots]=A(t_1)B(t_2)C(t_3) \dots,\nn \ee
for $t_1\ge t_2\ge t_3 \dots$. This prescription preserves the
quantum commutation relations (\ref{4}), (\ref{5}).

We define the rotation vector angle as,
\be\vec{\Omega}_i(t)=\sin^{-1}|\hat{z}\times\vec{M}_i(t)|\cdot\frac{\hat{z}
\times\vec{M}_i(t)}{|\hat{z}\times\vec{M}_i(t)|}\label{7}\ee
with $\vec{M}(\vec{r},t)$ being the staggered magnetization, where
for example, $\vec{Q}=0$ for ferromagnetic spin fluctuations. More
generally,
\be \vec{M}_i(t)=(-1)^{\vec{R}_i} \vec{S}_i (t). \ee

The transformed Hamiltonian is
\be\bar{H}(t)&=&\bar{H}_0(t)+\bar{H}_J(t)
\\\nn &=&U^\dag(t)[H_0(t)+ H_J(t)]U(t)\label{12}\ee
where $\bar{H}_0$ is given for a free electron band by
\be \nn \bar{H}_0(t)&=&-\frac{\hbar^2}{2m}\sum_{ss'}\int
\textnormal{d}
\vec{r}\:\psi_s^\dag(\vec{r},t)[\vec{\nabla}\delta_{s s'}+{}\\\nn
& & {}i A_{s s'}(t) ]^2\:\psi_{s'}(r,t). \ee
Here the gauge field $\vec{A}_{s s'}(\vec{r},t)$ is defined as
\be \vec{A}_{s s'}(\vec{r},t)=\frac{1}{2}\vec{\sigma}_{s
s'}\cdot\nabla\vec{\Omega}(r,t)\label{}.\ee
Note that $\bar{H}_0$ is just the electronic kinetic energy of
quantum electrodynamics (QED) for $e/c\equiv1$.

It is convenient to expand out $\bar{H}_0$ in powers of $\vec{A}$
and we find,
\be\bar{H}_0(t)=H_0(t)+H_{sdp}(t)+H_{dia}(t)  \label{}\ee
with $H_0(t)$ given by (\ref{2}). The two new spin couplings are
(i) the spin deformation potential (sdp) $H_{sdp}$,
\be
H_{sdp}(t)&=&\frac{i\hbar^2}{4m}\int\textnormal{d}\vec{r}\sum_{s
s'}[\psi_s^\dag(\vec{r},t)\:\vec{\sigma}_{s s'}{}\\\nn & &
{}\cdot\vec{\nabla}\psi_{s'}(r,t)][\vec{\nabla}\vec{\Omega}(\vec{r},t)+\vec{\Omega}(r,t)\vec{\nabla}],
\ee
where the spin current is defined by
\be \vec{j}_s(\vec{r},t)\equiv-\frac{i\hbar}{2}\sum_{s
 s'}\psi_s^\dag(\vec{r},t)\:
 \vec{\nabla}\psi_{s'}(\vec{r},t)\:\vec{\sigma}_{s s'} \ee
 $\vec{j}_s(\vec{r},t)$ couples to the quantity $\nabla\vec{\Omega}(\vec{r},t)$, in the same manner as the
$\vec{p}\cdot\vec{A}+
 \vec{A}\cdot\vec{p}$ paramagnetic coupling of QED.
 The \textquotedblleft diamagnetic" term proportional to $\vec{A}(\vec{r},t)^2$ is
 given by
 \be H_{dia}(t)&=&-\frac{\hbar^2}{8m}\int\textnormal{d}\vec{r}\sum_s\psi_s^\dag(\vec{r},t)
 \psi_s(\vec{r},t){}\\\nn & &
{}|\nabla\vec{\Omega}(\vec{r},t)|^2 \ee
 It is $H_{sdp}$ and $H_{dia}$ which lead to the anomalous exponents
 near the QCP.

 $\bar{H}_J(t)$ is given by
 \be H_J(t)=\frac{J}{2}\int\textnormal{d}\vec{r}\sum_{\bar{s}}\bar{s}\:\psi_{\bar{s}}^\dag
 \:\psi_{\bar{s}}(\vec{r},t)\:|S(\vec{r},t)|\:(-1)^r, \ee
where $\bar{s}$ is defined as the $\hat{z}$ in the spin rotated
frame. We note that the transformed exchange coupling acts as a
simple diagonal spin Zeeman coupling, analogous to the exchange
interaction in the mean field approximation. This exact result
greatly simplifies the calculations.

While the discussion at this point is exact, it is helpful to make
pairing correlations manifest by elevating $\psi_{i s}^\dag(t)$ to
a two component pseudo-spin operator,
\be \Psi_{i s}^\dag(t)=[\psi_{i\uparrow}^\dag(t),
\psi_{i\downarrow}(t)] .\ee
We introduce the Pauli pseudo-spin matrices $\tau_i$, with
$i=0,1,2,3$. Then $\bar{H}(t)$ can be written as
\be\bar{H}_0(t) &=&
-\frac{\hbar^2}{2m}\sum_{ss'}\int\textnormal{d}\vec{r}\:
\Psi_s^\dag(r,t)\:[\nabla\tau_{3}\delta_{ss'}+{}\\\nn & &
{}i\vec{A}_{ss'}(\vec{r},t)]^2\:\Psi_{s'}(\vec{r},t)\ee
and
\be\bar{H}_J(t) = J\sum_{\bar{s}}\Psi_{\bar{s}}^\dag(r,t)\openone
\Psi_{\bar{s}}\:S_{\hat{z}}(t),\ee
where $\bar{s}$ and $\hat{\bar{z}}$ correspond to the direction of
the local $\vec{M}(\vec{r},t)$. The one loop Gor'kov self energy
is given by
\be\label{bb}\not\!\Sigma(\vec{k},\omega_n)&=&{}\\\nn & &
\sum_{\vec{Q},\omega_m}[V(Q, \omega_n -\omega_m)\:\not\! G(k-Q,
\omega_n-\omega_m)] \ee
where $\omega_n=(2n+1)k_BT$, $\omega_m=2\pi mk_BT$ and $V$ is the
potential arising from $\bar{H}_{sdp}$ and $\bar{H}_{dia}$. Taking
the 1, 2 components of (\ref{bb}) we have the gap equation,
\be\not\!\Sigma(k,\omega_n)_{1 2}&=& {}\\\nn & &
\sum_{\vec{Q},\omega_n}[V(Q, \omega_n -\omega_m)\:\not\!G(k-Q,
\omega_n-\omega_m)]_{1 2}. \ee
The normal phase renormalization shifts are given by the
$\openone$ and $\tau_3$ terms in (\ref{bb}), respectively.

While $\bar{H}_{dia}$ and $\bar{H}_{J}$ give $s-$wave $(l=0)$ and
$d-$wave $(l=2)$ $(S=0)$ pairing, $T_C$ is exponentially larger
for $p-$wave, $(l=1)$, $(S=1)$ pairing. To calculate the effective
pairing interaction in the $p-$wave, $S=1$ channel we introduce
quasi particles which diagonalize $H_0(t)$ and $\bar{H}_J(t)$.
These terms in $H$ couple each $k$ with $k\pm Q$ in reciprocal
space, without spin flip. Introducing the Bogoliubov-Valatin
quasi-particle transformation, one has
\be\gamma_{k,s}=u_k\psi_{k s}+v_{k s}\psi_{k+Q s}\ee
and
\be\gamma_{k+Q,s}=v_{k s}\psi_{k s}-u_k\psi_{k+Q, s}\ee
with eigenvalues
\be\pm E_k=\pm\sqrt{\epsilon_k^2+(J/2)^2}\ee
Here,
\be u_{ks}\equiv\sqrt{1/2(1+\epsilon_k/E_k)},\\ \nn
v_{ks}\equiv\sqrt{1/2(1-\epsilon_k/E_k)}.\ee
In the $\gamma$ representation, $H_{sdp}(t)$ becomes
\be H_{sdp}(t)&=&-\frac{i\hbar^2}{4m}\sum_{s
s'}\gamma_s^\dag(\vec{r},t)\bold{\openone}\gamma_{s'}
(\vec{r},t)(\nabla\vec{\Omega}(\vec{r},t){}\\\nn & &
{}+\vec{\Omega}(\vec{r},t)\nabla)\ee
with
\be E_k=\sqrt{(\frac{\hbar^2k^2}{2m}-\mu)^2+\frac{J^2}{4}}\ee
for the free electron model.


The ground canonical potential is given by,
\be \Lambda(T) &=&
-k_BT\ln\textnormal{Tr}\:T\:e^{-\beta(\bar{H}-\mu N)}{}\\\nn
&=&-k_BT\ln Z(T) .\ee
Within the RPA for $\bar{H}_{sdp}(t)$ and $\bar{H}_{dia}(t)$, as
shown in Fig(1),one has,
\be \Lambda(T)_{RPA} &=& \sum_{\vec{Q},
\omega_m}\textnormal{Tr}\{V(\vec{Q},\omega_n)\Phi_0(Q,\omega_n){}\\\nn
& & {} [1-V(Q,\omega_n)\Phi_0(Q,\omega_n)]^{-1}\}, \ee
where the zero order irreducible polarizability is defined by
\be\Phi_0(\vec{Q},\omega_m)= \sum_{k,\omega_n}\:\not\!G_0(\vec{k}+
\vec{Q},\omega_n+\omega_m)\:\not\!G_0(\vec{k},\omega_n) \ee
The specific heat is given by
\be C_V(T)\equiv-\frac{\textnormal{d}}{\textnormal{d}T}\:k_BT\ln
Z(T). \ee
We find $C_V(T)\simeq\ln T$ for $W/100 < T <W$, as obsereved in
experiment.

The RPA spin susceptibility is given by
\be \chi_{RPA}(Q,\omega_n,T)&=&\frac{1}{4}\sum_{s \bar{s}}
s\bar{s}\Phi_0(Q,\omega_n,T){}\\\nn & &
{}[1-V(Q,\omega_n,T)\Phi_0(Q,\omega_n)]^{-1}\ee
%

To estimate $T_C$, we use a square potential model, as in the BCS
model. Then
\be T_C = 1.14\:\omega_s\:e^{-\frac{1+\lambda}{\lambda}}, \ee
where \be\lambda = N(0)V_{pairing},  \ee
and $V_{pairing}$ being given by (\ref{bb}).
%

To determine $T_C$ we take the limit that the magnitude of the
order parameter $\Delta(k,\omega_n,T)$ goes to zero. One finds
that $T_C$ satisfies the linear gap equation
\be\Delta(\vec{k},\omega_n,T_C)&=& {}
\\\nn & & {} -\sum_{Q,\omega_m}\frac{1}{Z(\vec{k},\vec{Q},\omega_m)}(u_kv_{k+Q}+ v_ku_{k_Q})^2{}\\\nn & &
{}V(\vec{k},\vec{Q},\omega_m,T)\:G_0\:(k+Q,\omega_n+\omega_m){}\\\nn
& & {}G_0(-\vec{k}-\vec{Q},-\omega_n-\omega_m)\:\Delta(\vec{k}
+\vec{Q},\omega_n+{}
\\\nn & & {}\omega_m,T)\label{tc1}\ee
where $Z(k,Q,\omega_m)/\omega_n$ is the parallel component of
$\not\!\Sigma$ (see (\ref{bb})). The quantity $(uv'+vu')$ is
nothing but the BCS coherence. For the BCS square potential in
energy, with a cut off at spin fluctuation frequency $\omega_s$,
\be V(\vec{k},\vec{Q},\omega_m,T)=V_0(\vec{k},Q,T),\quad\quad\quad
|\omega_m|\leq\omega_s,\ee
one finds
\be k_BT_C\simeq\omega_se^{-\frac{1+\lambda}{\lambda}},\ee
where the coupling constant $\lambda$ is given by
\be\lambda&=&
N(0)(\frac{\hbar^2}{4m})^2[\frac{|V_0(Q_0,T)|^2}{\omega_s} {}
\\\nn & & {}\langle(u_kv_{k+Q}+v_ku_{k+Q})^2\rangle].\ee
Using
\be \omega_s\simeq\frac{J^2}{W},\ee
we obtain
\be k_BT_C\simeq\frac{J^2}{W}e^{-\frac{1+\lambda}{\lambda}}\ee
Maximizing $T_C$ one finds
\be k_BT_C\simeq\frac{W}{8}\ee
for $J\simeq 0.1\:W$. For $W\simeq 10 eV$, this leads to the
remarkably high value of
\be k_BT_C\simeq 1 eV\simeq 10^4 K,\ee
with $J=1\:eV$. Clearly this result is of immense technological as
well as scientific value.

Details of this new theory with applications to electronic
tunneling, ARPES, transport measurement, magnetic penetration
depth, $C_V(T)$, etc. will soon be published elsewhere (PR).
\acknowledgments  The author wishes to thank Profs. Megan Aronson,
Nicholas Bonesteel, Lev P Gor'kov and Kun Yang for helpful
discussions. This work was supported in part by a grant from the
National Science Foundation, grant No. 0084173 and Department of
Energy, grant No. DE-FG03-03NA00066 The author would like to
express his sincere thanks to Layla Hormozi for help with
preparation of the manuscript.
\begin{figure}
\centerline{\includegraphics[height=2cm,angle=0]{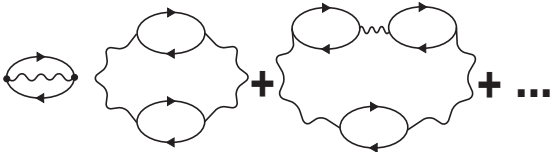}}
\caption{The grand canonical $\Lambda(T)$ within the random phase
approximation.} \label{fig1}
\end{figure}
\begin{figure}
\rightline{\includegraphics[clip,width=3.3in]{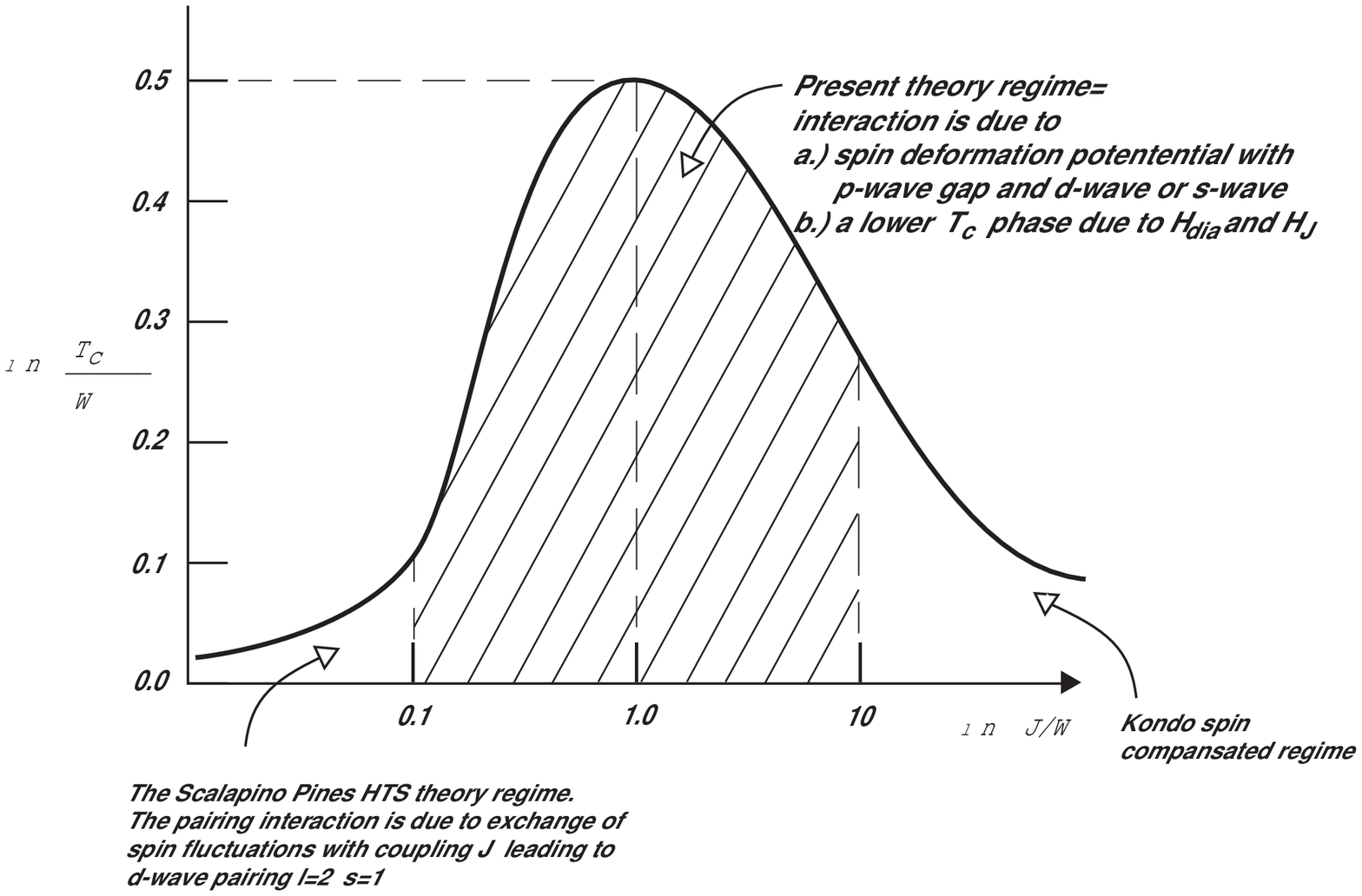}}
\caption{The phase diagram of the $t-J$ model, showing the
conventional nearly antiferromagnetic fermi liquid of Pines and
Scalapins for $J\leq 0.01\:W$, where $J$ is the electron localized
spin exchange coupling and $W$ is the electronic band width. For
$0.01\:W\leq J\leq 10\:W$ a novel $p-$wave, $l=1$, $s=1$ phase is
predicted with an extremely high $T_C$ of immense technological
importance (see the text). In this phase the existence of a
Leggett collective model is predicted, corresponding to an
oscillation at frequency $\omega_L$ of the angle between $\vec{L}$
and $\vec{s}$ vectors of a the pair. However, here the novel
strong spin deformation raises $\omega_L$ to a high value of order
the optical range $vs$ the low frequency of superfluid ${}^3He$,
where the spin orbit coupling $H_{so}$ is extremely weak.}
\label{fig2}
\end{figure}

\end{document}